\documentstyle[12pt]{article}
\newenvironment{namelist}[1]{%
\begin{list}{}
{

\settowidth{\labelwidth}{#1}
\setlength{\leftmargin}{\labelwidth}
}
}{%
\end{list}}
\newcommand{\dis}{\displaystyle}
\begin{document}
\def\qed{~\vrule height6pt width4pt depth0pt\medskip}
\setlength{\baselineskip}{0.35in}
\setlength{\jot}{0.2in}

\begin{center}
{\large \bf
On the possibility of a complex 4-dimensional space-time manifold
}\\

\vspace*{0.5cm}
Lu Lin
\\
Department of Electrophysics\\
National Chiao Tung University\\
Hsinchu, Taiwan \\
Republic of China\\
\end{center}

\vspace*{2cm}

\noindent
{\bf ABSTRACT}~~\\

The possibility of a complex 4-dimensional space-time
manifold is suggested.  This may imply the existence of a
matter wave.\\
\newpage
\noindent
1. Dirac theory:\\

To describe the motion of a free electron, Dirac (Dirac 1958) introduced
$(\alpha _{1}\alpha_{2}\alpha_{3}\beta )$ or equivalently
$(\gamma^{\mu }, \mu =1,2,3,4)$ (Bathe 1986)into the theory.  The $\gamma 's$
are independent of the $x_{\mu }'s$ and the $p's$, where
$x$ and $p$ are the sapce-time coordinates and the momentum-energy of the
particle.  Dirac pointed out that the $\gamma 's$ describe some new
degree of freedom, belonging to some internal motion in the electron, so
that they must commute with the $x's$ and the $p's$.  The
basis state vector is represented by two seperated pieces
of components, a space-time
part $(x_{\mu})$ and a spin part which is a complex spinor.  In
this theory, the quantity $j^{\mu}=\bar{\Psi}\gamma^{\mu} \psi$ is a
Lorentz 4-vector.  Since the $\gamma 's$ are independent of the $x's$
and the $p's$, so $j^{\mu }$ is also independent of the $x 's$ and
$p's$.\\

In mechanics, the infinitesmal generators for the $x 's$ are the
dynamical conjugete variables $p's$, and for rotational angular
displacements $\theta '_{i}s$ the angular momentum.  The infinitesmal
unitary transformation in
spin space can be written as
\begin{eqnarray}
U=1+\sum\frac{1}{2}\sigma_{i}\cdot \xi_{i} ,
\end{eqnarray}

\noindent
where the $\sigma 's$ are the Pauli matrices and the $\xi 's$ the
analogue quantities of the $\theta '_{i}s$ in spin-space.  The relations
between the $\sigma 's$ and the $\gamma 's$ are given as (Bethe 1986)
\begin{eqnarray}
& & (\gamma ^{\mu})=(\beta \alpha _{i}, \beta ), ~~~~(i=1,2,4)\\
& & \alpha _{i}=\rho_{1}\sigma _{i}, \nonumber
\end{eqnarray}

\noindent
where $\rho_{1}$ is given in (Dirac 1958).  The
$\gamma 's$ are linear combinations of the $\sigma '_{i}s$.
Consider the $\gamma 's$ as a set of dynamical variables
 $\left (\dis\frac{d}{dt}
\gamma \neq 0\right )$, their conjugate variables should have
$(\eta_{1}\eta_{2}\eta_{3})$ as the analogue of $(\theta_{1}\theta_{2}\theta_{3})$
plus a 4th-component $\eta_{4}$.  The $\theta 's$ are in the spin-space
so they are independent of the $x's$ and the $p's$, therefore
$\eta_{1},\eta_{2}$ and $\eta_{3}$ are also independent of the
$x's$ and the $p's$.  It is reasonable to assume that $\eta_{4}$ is
also independent of the $x's$ and the $p's$ since the $\gamma 's$
describe some new dagree of freedom which are internal in the particle, their
dynamical conjugate variables should also be related to variables
which are internal.\\

Now, we can see clearly that the basis state vector of the particle
is 8-dimensional.  However, nature should prefer simple and symmetry.
Instead of having two seperated parts of basis space, we should
rather have one whole complete piece of continuous basis space.
That is, it is natural to suggest a complex 4-dimensional space-time
manifold.\\

The above conerideration is based only on the dimensionality of Dirac
theory and does not depend on the content of the theory.  In fact,
we can use Schr$\ddot{\rm o}$dinger-Pauli equation (Dirac 1958) where the
basic variables are $(xyzt\sigma_{1}\sigma_{2}\sigma_{3})$, where the
$\sigma 's$ operate on the spin-space.  The infinitesmal unitary
transformation in spin space has the form as equation (1) and thus we
have three conjugated variables similar to $(\xi_{1}\xi_{2}\xi_{3})$.  Now, if
we wish the theory to be Lorentz convariant, we have to generalize the three
$\sigma 's$ to a 4-component quantity which is similar to the $\gamma 's$.
Therefore we again have four more variables besides $(xyzt)$.\\

\noindent
2. Matter wave:
Consider a complex 4-dimensional space-time manifold, to describe the motion
of a free particle,
there must exist a well-behaved function $\Psi (T=t+i\bar{t},
X=x+i\bar{x})$ which is continuous and differatiable in the domain under
consideration.  The simplest form of such a function can be written as
\begin{eqnarray}
& & \Psi =\Psi (T)\cdot \Psi (X),\\
& & T=t+i\bar{t},\\
& & X=x+i\bar{x}.
\end{eqnarray}

\noindent
$\Psi (T)$ and $\Psi (X)$ must satisfy the Cauchy-Riemann conditions
(Churchill 1948) for an analytic function.  Write $\Psi (X)$ as
\begin{eqnarray}
\Psi (X)=u(x, \bar{x})+iv(x,\bar{x}).
\end{eqnarray}

\noindent
Since $\Psi (X)$ is analytic in some region of the $X$ plane, we
have
\begin{eqnarray}
& & \frac{\partial ^{2}u}{\partial x^{2}}+
\frac{\partial ^{2}u}{\partial \bar{x}^{2}}=0,\nonumber\\
& & \frac{\partial ^{2}v}{\partial x^{2}}+\frac{\partial ^{2}v}{\partial \bar{x}^{2}}
=0.
\end{eqnarray}

\noindent
Then we can also write
\begin{eqnarray}
c_{1}\left ( \frac{\partial ^{2}u}{\partial x^{2}}
+\frac{\partial ^{2}u}{\partial \bar{x}^{2}}\right )
+c_{2}\left ( \frac{\partial ^{2}v}{\partial x^{2}}
+\frac{\partial ^{2}v}{\partial \bar{x}^{2}}\right )=0,
\end{eqnarray}

\noindent
where $c_{1}$ and $c_{2}$ are constants.  Assume that nature does not have
special preference between real and imaginery parts except the $i$.
Take $c_{1}=1$ and $c_{2}=i$, we get
\begin{eqnarray}
\frac{\partial ^{2}\Psi }{\partial x^{2}}
+\frac{\partial ^{2}\Psi }{\partial \bar{x}^{2}}=0.
\end{eqnarray}

\noindent
Since $x$ and $\bar{x}$ are basically independent, we must be able to
describe the $x$-part  and the $\bar{x}$-part of the function seperately.
That is, we have $\Psi (x,i\bar{x})=\Psi (x)\Psi (i\bar{x})$.  Then
the
solution of equation (9) has the form
\begin{eqnarray}
\Psi (X)=Ae^{\pm ax}\cdot e^{\pm b\bar{x}}, ~~~~~a^{2}+b^{2}=0.
\end{eqnarray}

\noindent
where $A,a$ and $b$ are constants.
 We take $\Psi (X)=Ae^{\pm ikX}=Ae^{\pm ik(x+i\bar{x})}$.  Such that
$\Psi$ is analytic.\\
Similarly we find $\Psi (T)$ to be $e^{\pm i\omega T}$, therefore
we have
\begin{eqnarray}
\Psi (X,T)=Ae^{\pm i(kX-\omega t)}.
\end{eqnarray}

With respect to our direct experimental meaurements, we can take only real
space-time variables.  The projection of $\Psi (X, T)$ onto the real
space-time region gives us $\Psi (x,t)=Ae^{i(kx-\omega t)}$, the
constants $k$ and $\omega $ are identified as the wave number and
angular frequency.  Up to this point, we can see that the free particle
we wish to describe is a free wave.  This may offer a possible solution
to the question that why a particle has to be a wave as well.\\

Finally, if the space-time is complex, there may have a possibility that those
physically undetectable particles may stay in the imaginery part of
the nature.\\

The author wishes to express his appreciation to Professors E. Yen
J.C. Shaw, J.C. Lee, B. Rosenstein, C.S. Han, T.H. Kou, C.S. Chu and W.D. Nee
 for valuable discussions.\\

\noindent
References
\begin{namelist}{Churchill R V 1948}
\item [{Dirac P A M 1958}]
The principles of quantum mechanics, fourth ed. Oxford Univ. Press, London.
\item [{Bethe H A Jackiw R W 1986}]
Intermediate quantum mechanics, third ed.,
Addison-Wesley Publishing Co. New York
\item [{Churchill R V 1948}]
Introduction to complex variables and application, McGraw-Hill Book
Co. Inc. U.S.A.
\end{namelist}
\end{document}